\documentclass[11pt, a4paper, onecolumn, copyright]{AweAI}

\usepackage{multirow}
\usepackage[authoryear, sort&compress, round]{natbib}
\usepackage{float}
\usepackage{graphicx}
\usepackage{xspace} 

\usepackage[table]{xcolor} 
\usepackage{tcolorbox}
\usepackage{listings} 

\tcbuselibrary{listings, skins, breakable}

\newtcblisting{markdownbox}[1][]{
  listing engine=listings,
  listing options={
    basicstyle=\small\ttfamily,
    breaklines=true,
    breakatwhitespace=true,
    numbers=none,
    frame=none, 
    columns=fullflexible,
    showstringspaces=false,
    xleftmargin=0pt,
    xrightmargin=0pt,
    aboveskip=0pt,
    belowskip=0pt
  },
  colback=gray!5!white,
  colframe=gray!70!black,
  title={\textbf{Prompt (Markdown)}},
  fonttitle=\bfseries,
  listing only,
  left=5mm, right=5mm, top=3mm, bottom=3mm,
  breakable,
  enhanced,
  overlay={\begin{tcbclipinterior}\fill[gray!5!white] (frame.south west) rectangle ([xshift=0mm]frame.north west);\end{tcbclipinterior}},
  #1
}
\definecolor{bg}{rgb}{0.95,0.95,0.95}
\uselogo{}

\DeclareRobustCommand{\github}{\raisebox{-1.5pt}{\includegraphics[height=1.05em]{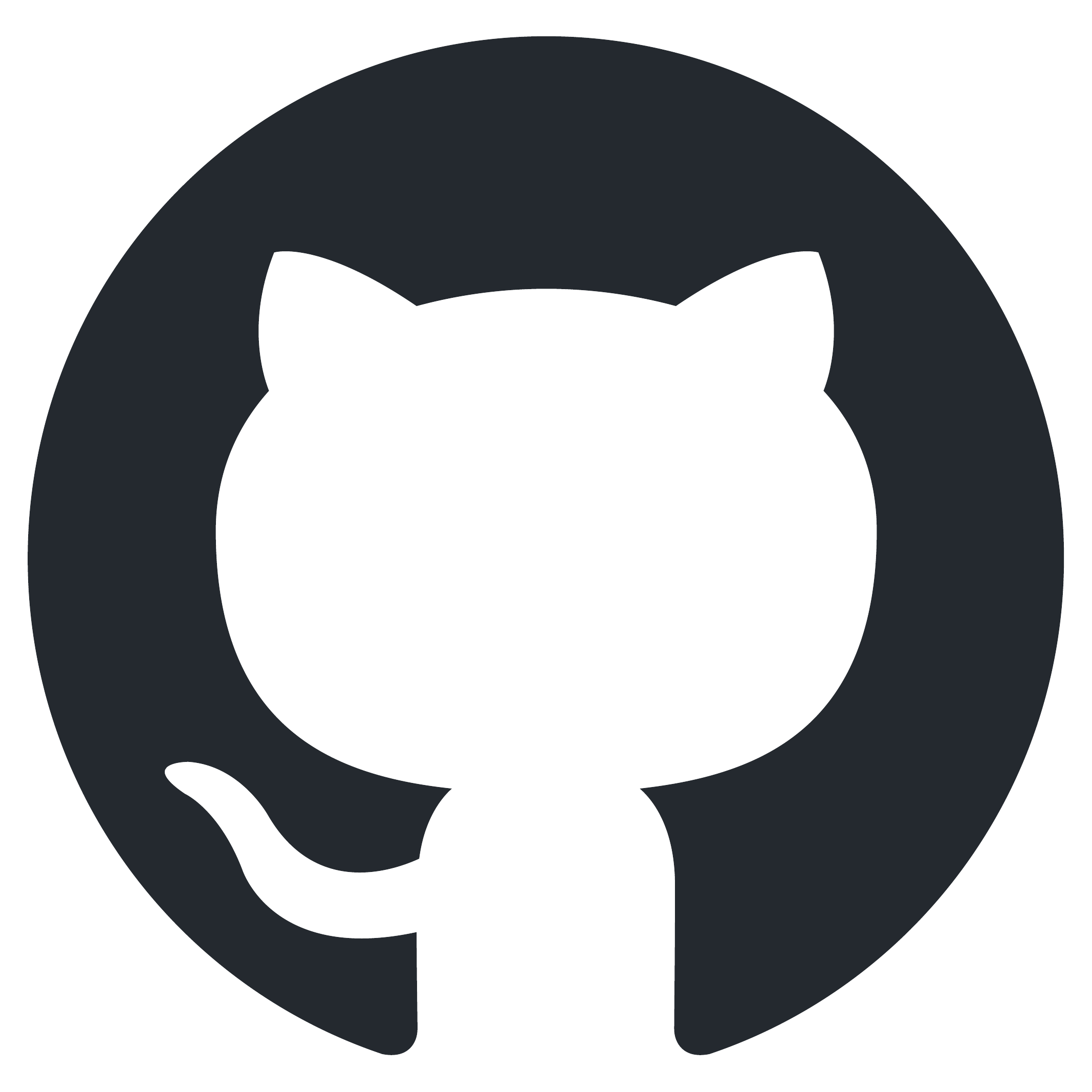}}\xspace}
\DeclareRobustCommand{\huggingface}{\raisebox{-1.5pt}{\includegraphics[height=1.05em]{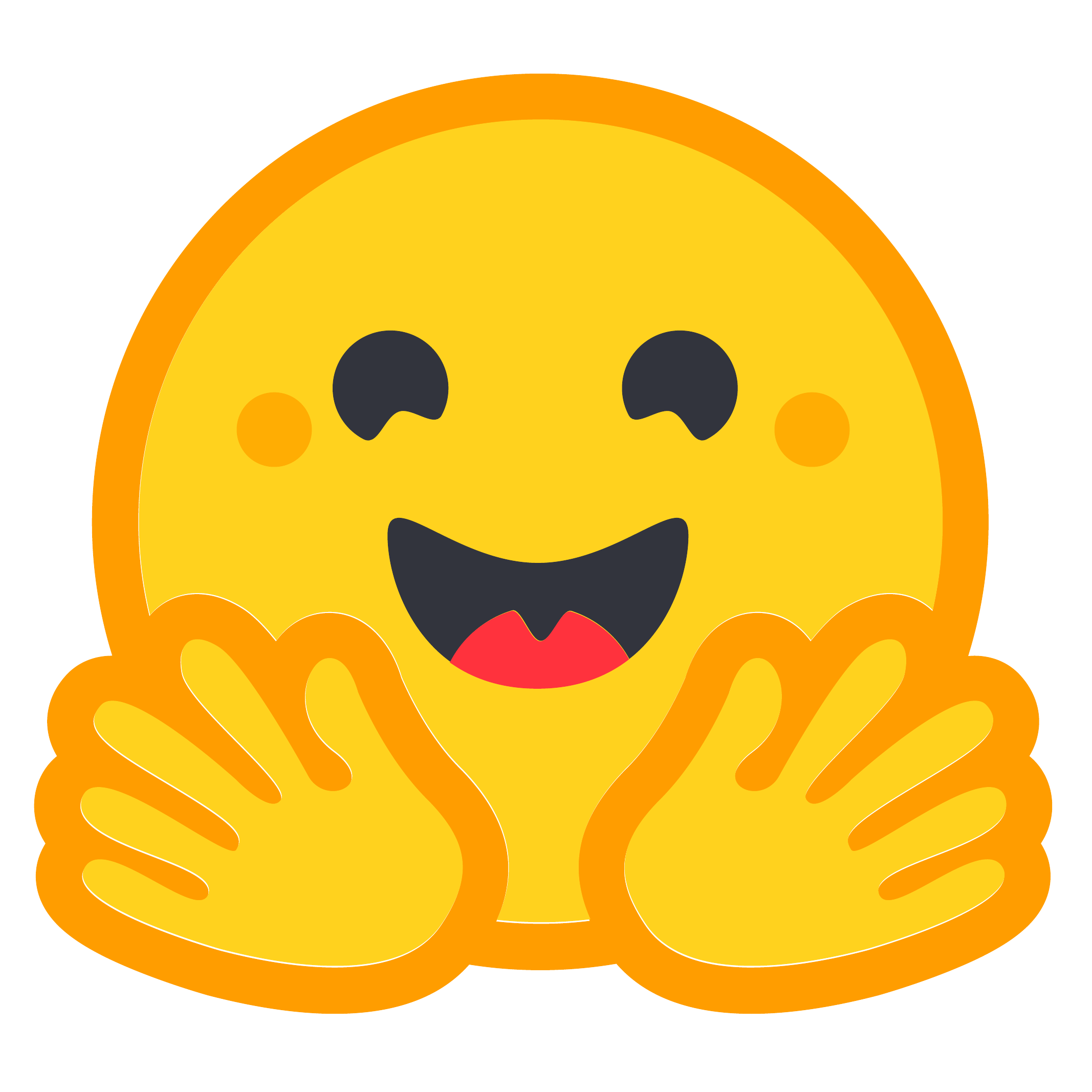}}\xspace}
\DeclareRobustCommand{\hfdataset}{\raisebox{-1.5pt}{\includegraphics[height=1.05em]{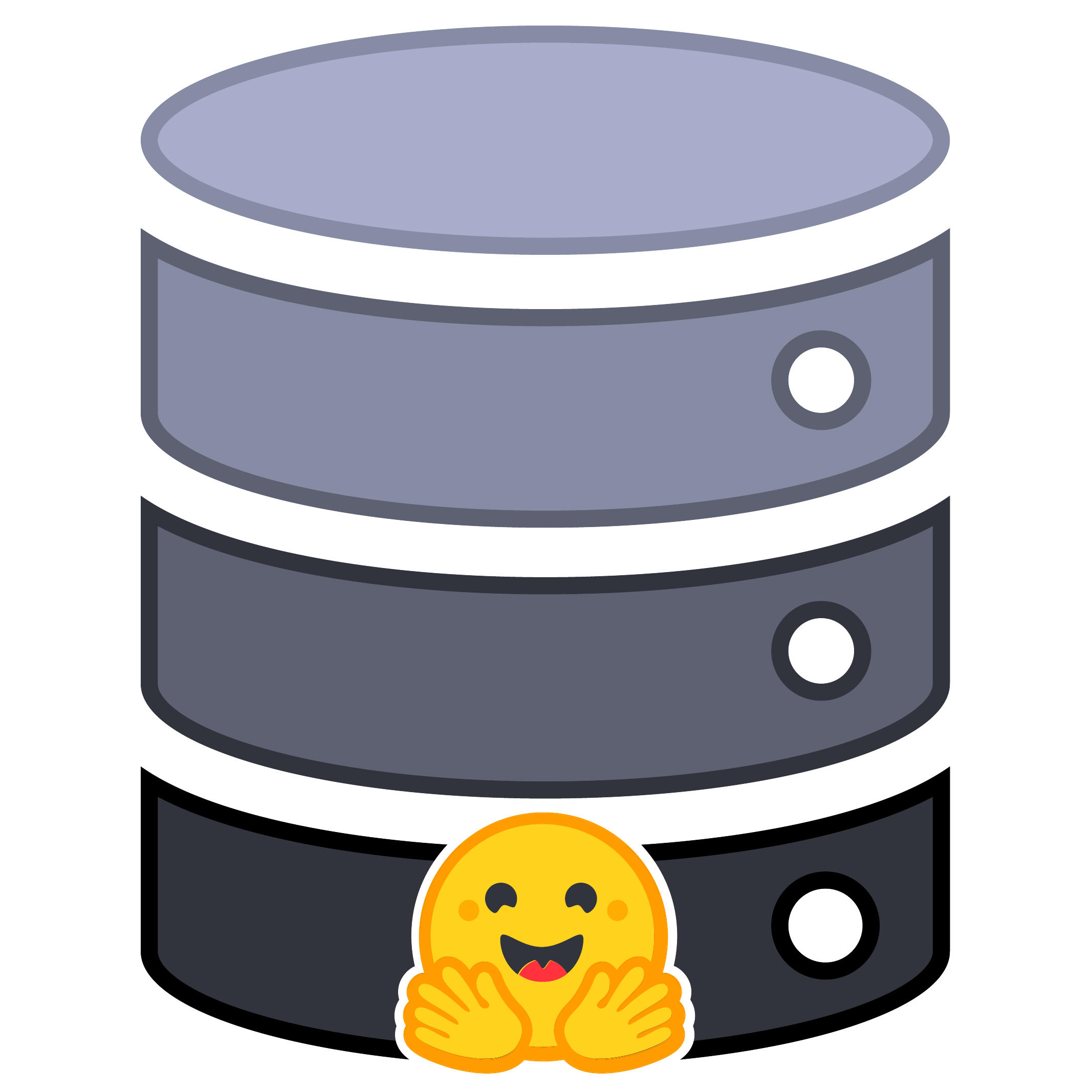}}\xspace}

\newcommand{\footerlinks}{%
  \huggingface\ \href{https://huggingface.co/AweAI-Team/Scale-SWE-Agent}{Model}\hspace{1.5em}%
  \hfdataset\ \href{https://huggingface.co/collections/AweAI-Team/scale-swe}{Dataset}\hspace{1.5em}%
  \github\ \href{https://github.com/AweAI-Team/ScaleSWE}{Code}%
}

\title{Immersion in the GitHub Universe: Scaling Coding Agents to Mastery}

\renewcommand{\today}{}
\newcommand{\publicday}{Feb.~10, 2026}

\author[*]{Jiale Zhao}
\author[*]{Guoxin Chen}
\author[*]{Fanzhe Meng}

\author[ \hspace{-0.3em}]{Minghao Li}
\author[ \hspace{-0.3em}]{Jie Chen}
\author[ \hspace{-0.3em}]{Hui Xu}
\author[ \hspace{-0.3em}]{Yongshuai Sun}
\author[ \hspace{-0.3em}$^\dag$]{Wayne Xin Zhao}
\author[ \hspace{-0.3em}$^\dag$]{Ruihua Song}
\author[ \hspace{-0.3em}]{Yuan Zhang}
\author[ \hspace{-0.3em}]{Peng Wang}
\author[ \hspace{-0.3em}]{Cheng Chen}
\author[ \hspace{-0.3em}]{Ji-Rong Wen}
\author[ \hspace{-0.3em}$^\dag$]{Kai Jia}

\correspondingauthor{\{marshmallowzjl, gx.chen.chn, mengfanzhe16, batmanfly\}@gmail.com, songruihua\_bloon@outlook.com, jiakai@bytedance.com}

\affil[1]{Gaoling School of Artificial Intelligence, Renmin University of China}
\affil[2]{BandAI, ByteDance}
\affil[3]{AweAI Team\footnote{$^*$Equal Contributions. $^\dag$Corresponding authors.\hfill\textbf{Date:} \publicday.}}

\begin{abstract}
Achieving mastery in real-world software engineering tasks is fundamentally bottlenecked by the scarcity of large-scale, high-quality training data. Scaling such data has been limited by the complexity of environment setup, unit-test generation, and problem statement curation. 
In this paper, we propose \textbf{Scale‑SWE}, an automated, sandboxed multi‑agent workflow designed to construct high‑quality SWE data at scale. The system coordinates three specialized agents—for environment setup, test creation, and problem description synthesis—to process 6 million pull requests across 5.2k repositories, producing \textbf{Scale‑SWE‑Data}: 100k verified SWE instances, the largest such dataset to date. It substantially surpasses existing real‑world datasets in repository diversity and reflects realistic task complexity.
We further demonstrate the dataset’s utility for training by distilling 71,498 high‑quality trajectories and fine‑tuning Qwen‑30B-A3B-Instruct to produce \textbf{Scale‑SWE‑Agent}. Our agent achieves a 64\% resolve rate on SWE‑Bench‑Verified—a nearly three‑fold improvement over the base model. Scale‑SWE provides a scalable, reproducible approach for data construction to advance LLM‑based software engineering. 

\centerline{\footerlinks}
\end{abstract}

\makeatletter
\begin{document}

\begingroup
\makeatletter
\renewcommand{\thefootnote}{}
\renewcommand{\@makefnmark}{}
\maketitle
\makeatother
\endgroup

\section{Introduction}

\begin{figure}[H] 
  \centering 
  \includegraphics[width=0.8\columnwidth]{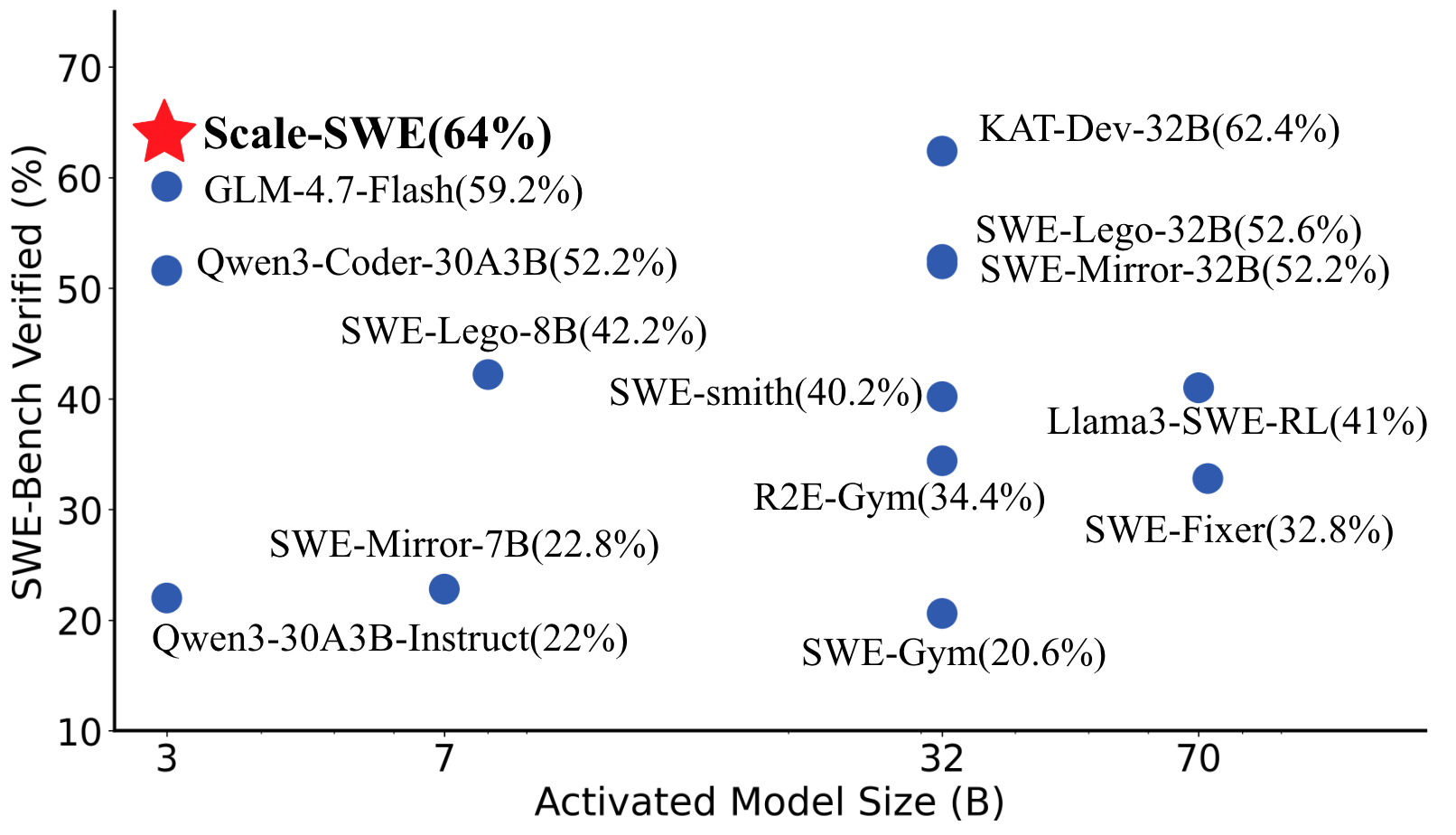} 
  \vspace{-0.3cm} 
  \caption{Resolved rate vs. activated model size on SWE-bench Verified. The vertical axis denotes the percentage of resolved issues on the SWE-bench Verified benchmark. The horizontal axis represents the number of activated parameters in billions (B).}
  \label{fig:performance}
  \vspace{-0.6cm} 
\end{figure}

Recently, LLM-based code agents have garnered significant attention for their demonstrated potential in tackling complex software engineering (SWE) tasks~\citep{anthropic2025claudesonnet45, googledeepmind2026gemini3pro, openai2025gpt52, team2025tongyi, chen2025iterresearch}, as reflected in benchmarks like SWE-bench~\citep{jimenez2023swe} and its successors~\citep{zhang2025swe}. Yet the advancement of these agents is fundamentally constrained by the scarcity of high-quality training data. Unlike conventional code generation, SWE tasks necessitate operation within executable environments, requiring agents to navigate existing codebases, manage dependencies, and satisfy test suites. These inherent complexities render the systematic curation and validation of appropriate data a significant challenge.

Current methodologies for constructing SWE-style datasets predominantly rely on labor-intensive manual curation~\citep{pan2024training} or on simplistic LLM-based~\citep{badertdinov2025swe} and rule-based synthesis~\citep{guo2025swefactory}. Consequently, existing datasets are often limited in scale, diversity, and difficulty, or lack the executable environments and comprehensive test suites. This limitation persists despite the wealth of available real-world software artifacts, including code repositories, issue trackers, and commit histories, which remain largely untapped for building a scalable, realistic dataset. The absence of systematic, automated mining techniques has thus produced a clear disconnect between these raw software resources and the creation of robust, large-scale training data. This compelling need drives our work to develop an automated and reproducible approach for dataset construction.

However, the automatic construction of SWE datasets presents unique and significant challenges. First, environment configuration becomes a major hurdle as repository diversity increases, leading to highly heterogeneous and often fragile build processes~\citep{froger2025scaling}, which can be challenging even for experienced developers to set up correctly. Second, real-world repositories frequently lack sufficient, well-defined unit tests. While incorporating these repositories is essential to reduce data bias and achieve scale, generating comprehensive unit tests itself is a complex problem that demands interactive agentic execution and self-correction.
Finally, a substantial portion of real-world pull requests either lack informative descriptions or are inherently unsuitable for SWE tasks. Therefore, generating high-quality, self-contained problem descriptions is challenging and requires agents to infer task intent by acquiring deep repository context through iterative sandbox exploration.

To overcome these challenges, in this paper, we introduce \textbf{Scale-SWE}, an automated, sandboxed multi-agent workflow designed for scalable, high-quality software engineering dataset construction. Our system coordinates three specialized agents: an \textit{environment builder agent} that sets up isolated Docker environments, a \textit{unit-test creator agent} that generates robust Pass-to-Pass (P2P) and Fail-to-Pass (F2P) test cases, and a \textit{problem statement agent} that crafts self-contained task descriptions grounded in pull request content. By processing 6 million pull requests across 5,200 repositories, this workflow produces 100,000 verified instances, yielding the largest SWE dataset to date, which we call Scale-SWE-Data. This dataset surpasses prior real-world datasets in repository diversity and reflects realistic software engineering complexity, both in the number of file modifications required and the robustness of its unit tests.

To further demonstrate the utility of Scale-SWE-Data for model training, we distill 71,498 high-quality trajectories from a subset of 25,000 instances using DeepSeek-V3.2. Fine-tuning Qwen3-30A3B-Instruct on this distilled data yields our Scale-SWE-Agent, which achieves a substantial performance boost on SWE-Bench-Verified, increasing the resolve rate from 22\% to 60\%. This result underscores the effectiveness of our dataset for training and enhancing LLM-based code agents. 

To summarize, our contributions are as follows:

\begin{itemize}[leftmargin=*, noitemsep, topsep=0pt]
    \item We introduce \textbf{Scale-SWE}, an automated, sandboxed multi-agent workflow for scalable, high-quality software engineering dataset construction. It systematically coordinates three specialized agents for environment setup, unit-test generation, and problem description synthesis.
    \item We construct \textbf{Scale-SWE-Data}, the largest verified SWE dataset to date, comprising 100,000 real-world instances. It surpasses prior benchmarks in repository diversity and task complexity, supporting both evaluation and training for LLM-based code agents.
    \item We distill 71,498 high-quality trajectories from Scale-SWE-Data and fine-tune Qwen-30A3B-Instruct to create \textbf{Scale-SWE-Agent}. The agent substantially boosts performance on SWE-Bench-Verified, achieving a 64\% resolve rate—a nearly three-fold improvement over the original backbone.
\end{itemize}

\section{Scale-SWE: Software Task Scaling}
\begin{figure*}[t!] 
  \begin{center}
    \centerline{\includegraphics[width=\textwidth]{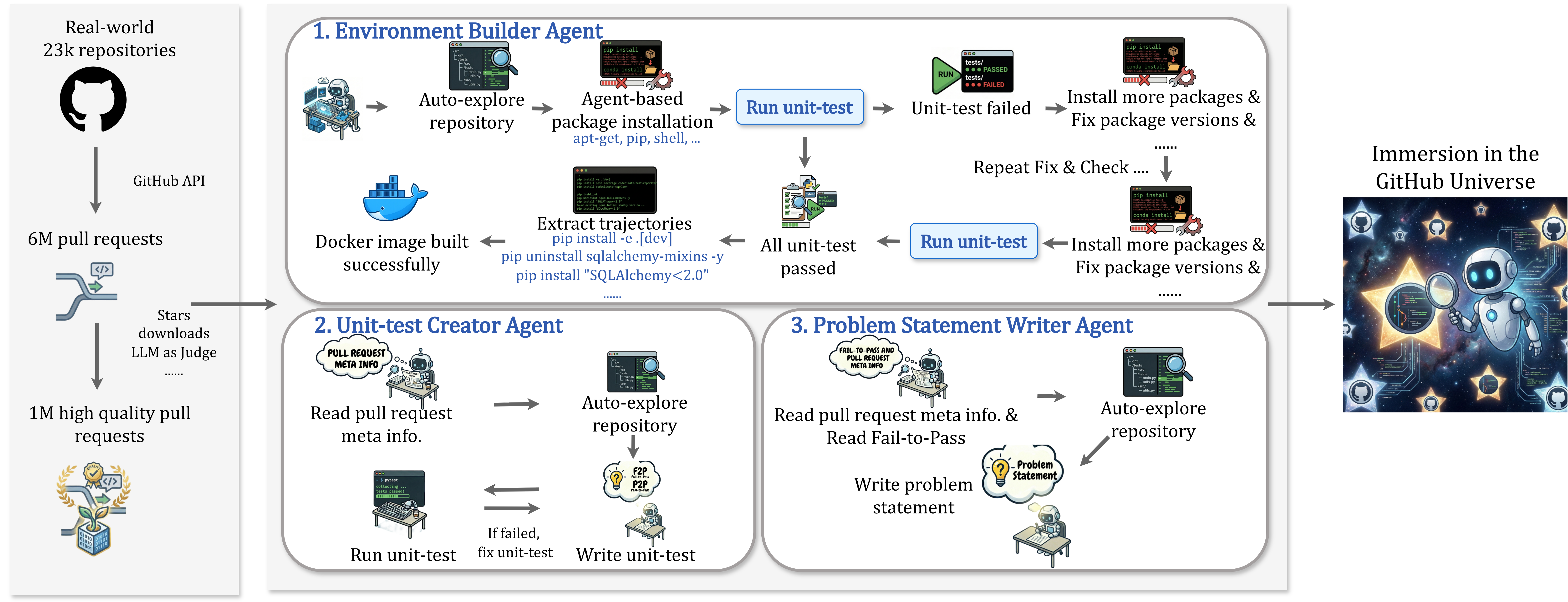}} 
    \caption{
The Sandboxed multi-agent system for Scale-SWE dataset construction. Starting from millions of raw GitHub pull requests, the pipeline employs a series of autonomous agents to transform high-quality PRs into executable software engineering tasks. The framework automates environment setup, unit test generation (Fail-to-Pass/Pass-to-Pass), and formal problem statement synthesis, ensuring the scalability and reproducibility of the distilled trajectories.
    }
    \label{system}
  \end{center}
  \vspace{-0.6cm} 
\end{figure*}

The core philosophy of Scale-SWE is to make use of a sandboxed multi-agent system to autonomously explore codebases and complete SWE data construction. Each SWE data instance encapsulates the necessary components: a docker image, a problem statement, and the validation unit tests (consisting of F2P and P2P unit tests). Our sandboxed multi-agent system offers significantly greater flexibility compared to prior rule-based construction methods. By shifting the construction burden to the agent, Scale-SWE enables the scaling of interactive environments while minimizing the heuristic bias inherent in rigid filtering pipelines.
In what follows, we firstly introduce our sandboxed multi-agent system (Section~\ref{sec:multi_agent_system}), and then detail by the data processing pipeline (Section~\ref{sec:data_process}).


\subsection{Sandboxed Multi-agent System}\label{sec:multi_agent_system}
To scale the automated construction of software engineering tasks, we develop a sandboxed multi-agent system. This framework leverages collaborative agents to produce a large volume of SWE data with fully integrated, executable environments.


\subsubsection{The Overall Workflow}
As illustrated in Figure~\ref{system}, our sandboxed multi-agent system automates the construction of software engineering tasks through the coordinated execution of three specialized agents: the Environment Builder Agent (EBA), the Unit-test Creator Agent (UCA), and the Problem Statement Writer Agent (PSWA). Within this framework, the {EBA} generates reproducible, Docker-based execution environments from target repositories, providing the isolated and consistent runtime needed for scalable task validation. The {UCA} synthesizes comprehensive, executable test suites—including both Fail-to-Pass (F2P) and Pass-to-Pass (P2P) test cases—from pull requests and repository context, ensuring robust evaluation criteria. Finally, the {PSWA} produces high-quality, self-contained problem descriptions grounded in the executable test suites, guaranteeing semantic alignment between the problem statement and the validation requirements. The system is built upon SWE-agent~\citep{yang2024sweagent} and powered by the DeepSeek language model family~\citep{liu2025deepseek} and Gemini3-Pro~\citep{googledeepmind2026gemini3pro}, with task instances generated using both DeepSeek v3.1 or DeepSeek v3.2, with the exception of PSWA, which utilizes Gemini3-Pro.

\subsubsection{Environment builder agent} 


The EBA is designed to automatically generate a reproducible, Docker-based execution environment from a target source code repository. By providing an isolated and executable sandbox, the EBA enables the scalable construction and validation of test samples for automated software engineering workflows.

\textbf{Role Formulation.} We define the core function of the EBA as the transformation of an initial, generic Docker environment into a specialized runtime container tailored to a specific repository. This process can be expressed as: 
\begin{equation}
    \mathcal{D}_{\text{final}} = \text{EBA}(\mathcal{R}, \mathcal{D}_{\text{init}}), 
\end{equation}
where \(\mathcal{R}\) denotes the input repository that the agent analyzes to infer dependencies and configuration logic, \(\mathcal{D}_{\text{init}}\) is the base Docker image, and \(\mathcal{D}_{\text{final}}\) is the resulting functional, ready-to-use container image.

\textbf{Construction Process.}
In implementation, the EBA is initialized within a base Docker container containing the cloned repository. It begins by autonomously exploring the repository's structure and analyzing configuration files—such as \texttt{setup.py}, \texttt{pyproject.toml}, and \texttt{README.md}—to infer project dependencies. Since Python projects lack a universal setup protocol, a standardized installation approach is insufficient. The agent must therefore interpret project-specific documentation and interactively resolve dependency conflicts by parsing terminal feedback. This feedback-driven, autonomous process enables flexible environment configuration, circumventing the limitations of static, rule-based methods (e.g., predefined \texttt{pip install} commands). Finally, we extract all executed commands from the agent's trajectory and use an LLM to synthesize them into a reproducible Dockerfile.


\textbf{Efficiently Scaling Environmental Diversity.}
While constructing a dedicated environment per pull request (PR) is straightforward, it is computationally prohibitive for large-scale evaluation. To scale across a diverse set of repositories, we sample at most ten PRs per repository for full environment construction via the EBA. This sampling strategy enables broader repository coverage while controlling resource costs. Note that ten PRs do not yield only ten task samples; multiple PRs can share the same environment if they originate from a similar runtime state, resulting in significantly more test samples than uniquely built environments. Specifically, for the remaining PRs, we execute tests in the ``nearest'' available Docker environment, determined by proximity in PR ID (as a proxy for repository timeline). If a PR fails the unit tests within its nearest available environment, it is discarded from the final dataset. Otherwise, the PR and its associated pre-established environment are retained as a valid test instance.  
On average, each repository contributes 19 test instances (see Table~\ref{tab:dataset_comparison}), dramatically improving environment reuse and overall dataset diversity.

\subsubsection{Unit-test creator agent} 
The \emph{unit-test creator agent (UCA)} is designed to automatically generate comprehensive and executable test suites from target source code repositories and their associated pull requests. By synthesizing semantic PR information with dynamic repository context, the UCA enables the scalable construction of validated Fail-to-Pass (F2P) and Pass-to-Pass (P2P) test cases, which are essential for robust software engineering task evaluation.

\textbf{Role Formulation.} The core function of the UCA is to transform the provided pull request metadata and its associated code context into a comprehensive suite of executable unit tests. This process can be expressed as:

\begin{equation}
    \mathcal{U} = \text{UCA}(\mathcal{M}, \mathcal{R}, \mathcal{D}_{\text{final}}),
\end{equation}
where \(\mathcal{M}\) denotes the input pull request metadata—including the title, description, and diff patches—that provides the semantic specification for test generation, \(\mathcal{R}\) is the associated source repository that the agent analyzes to understand code structure and logic, \(\mathcal{D}_{\text{final}}\) is the functional Docker environment built by the EBA, and \(\mathcal{U}\) is the resulting set of executable test cases. In this work, we adopt the SWE-bench protocol for unit tests, categorizing them into two types: Fail-to-Pass (F2P) tests, which initially fail on the original codebase and must pass after the bug-fixing patch to verify correctness; and Pass-to-Pass (P2P) tests, which are pre-existing passing tests that must continue to pass to ensure no regression is introduced.


\textbf{Construction Process.} 
A large proportion of GitHub repositories lack comprehensive unit tests, making the automated construction of both Fail-to-Pass (F2P) and Pass-to-Pass (P2P) test cases essential for creating valid Software Engineering (SWE) task instances.  
The UCA begins by analyzing structured metadata from a target PR, including its title, description, and diff patches. This input provides the necessary semantic grounding for the agent to comprehend the intent and scope of the proposed code changes. Building on this foundation, the agent performs an autonomous traversal of the associated repository to map its directory structure, identify key modules, and infer the underlying program logic and dependencies. 
However, generating effective unit tests presents a complex, stateful challenge that requires not only static code understanding but also dynamic behavioral validation. The agent must reason about cross-file interactions, data flow, exception handling, and edge cases—tasks that are difficult to accomplish through static analysis alone.  
To address this, the UCA is deployed within a secure, sandboxed execution environment (specifically, the Docker container $\mathcal{D}_{\text{final}}$ produced by the EBA). This sandbox grants the agent direct, real-time code execution privileges, enabling an interactive \textit{execute-analyze-refine} loop. The agent can thus dynamically run its proposed tests, observe their outcomes, and iteratively revise the test logic, assertions, and fixtures. This closed-loop, feedback-driven methodology allows the UCA to produce robust, executable test suites. 

\subsubsection{Problem Statement Writer Agent}

The \emph{problem statement writer agent (PSWA)} is designed to automatically synthesize high-quality, self-contained task descriptions from raw pull requests and their associated test suites. By grounding the narrative in executable unit tests, the PSWA ensures semantic alignment between the problem specification and unit tests, thereby producing well-posed and tractable software engineering tasks.

\textbf{Role Formulation.} The core function of the PSWA is to generate a formal problem statement free of solution leakage according to pull request metadata and its corresponding test suite. This process can be expressed as:
\begin{equation}
    \mathcal{S} = \text{PSWA}(\mathcal{M}, \mathcal{U}, \mathcal{R}, \mathcal{D}_{\text{final}}),
\end{equation}
where \(\mathcal{M}\) denotes the input PR metadata—including its title, description, and diff patches—that provides initial contextual grounding; \(\mathcal{U}\) is the executable unit-test suite generated by the UCA, which ensures the problem statement aligns with the actual validation requirements; \(\mathcal{R}\) is the associated source repository; \(\mathcal{D}_{\text{final}}\) is the functional Docker environment built by the EBA; and \(\mathcal{S}\) is the resulting formal problem description that articulates the issue without exposing implementation details and consistent with unit-test at the same time.

\textbf{Construction Process.}
Relying solely on raw PR descriptions as problem statements is fundamentally flawed due to their retrospective nature—they are often written after the fix is implemented and may leak solution details or reference internal artifacts. Moreover, a significant portion of high-quality PRs lack descriptions or are disconnected from original issue reports.
To overcome these limitations, the PSWA is initialized with both the PR metadata and the executable unit tests produced by the UCA. Integrating the test suite into the prompt is critical because F2P tests may invoke functions or classes that do not exist in the original codebase; the generated problem statement must explicitly articulate these requirements to make the task tractable. Without this alignment, descriptions derived only from PR metadata often diverge from the actual failures captured by the tests.
The agent thus synthesizes a coherent, self-contained narrative that specifies the expected behavior, any new interfaces required by the tests, and the context necessary for an external solver—all while deliberately omitting hints about the implementation. This process ensures that each synthesized task instance is both semantically precise and evaluation-ready, scaling the creation of SWE-bench-style problems from raw repository data.

For PSWA, we employ Gemini3-Pro, as our experiments indicate that it generates more consistent and rigorous problem statements while significantly minimizing information leakage.

\begin{table}[h]
    \centering
    \caption{Detailed statistics of the Scale-SWE dataset. We report the mean and percentiles (P50, P75, P95) for code modification metrics and test case distributions.}
    \label{tab:dataset_stat}
    \vspace{5pt}
    \begin{tabular}{lrrrrr}
        \toprule
        \textbf{Metric} & \textbf{Mean} & \textbf{P50} & \textbf{P75} & \textbf{P95} \\
        \midrule
        Modified Files & 6.4 & 3.0 & 6.0 & 18.0 \\
        \midrule
        Deleted Lines & 54.9 & 1.0 & 10.0 & 119.0 \\
        Added Lines & 220.8 & 43.0 & 120.0 & 595.0 \\
        Edited Lines & 37.0 & 6.0 & 20.0 & 108.0 \\
        Total Changes & 312.7 & 63.0 & 167.0 & 867.0 \\
        \midrule
        Fail-to-Pass & 5.7 & 2.0 & 5.0 & 15.0 \\
        Pass-to-Pass & 209.0 & 68.0 & 178.0 & 793.0 \\
        Total tests & 214.7 & 72.0 & 185.0 & 801.7 \\
        \bottomrule
    \end{tabular}
\end{table}

\subsection{Data Processing}\label{sec:data_process}
Ensuring high data quality is the foremost priority in building a reliable dataset. This section details our curation methodology to achieve this goal.

\textbf{Repository Selection.}
To assemble a comprehensive and high-quality corpus, we sourced repositories from two primary channels, illustrated in Figure \ref{workflow_overview}. First, we identified the repositories corresponding to the top 15,000 most-downloaded packages on the Python Package Index (PyPI), using data from \href{https://hugovk.github.io/top-pypi-packages/}{Top PyPI Packages}. Second, to capture notable projects beyond PyPI, we queried the \href{https://seart-ghs.si.usi.ch/}{SEART} search engine with the following criteria: primary language as Python, a minimum of 5 contributors, at least 500 stars, and a creation date between January 1, 2015, and October 29, 2025. This query returned 9,062 repositories. 
The initial union of these two sources yielded approximately 23,000 candidate repositories. We then applied a multi-stage filtering pipeline. First, to prevent data leakage from the evaluation benchmark, we excluded all PRs originating from repositories listed in SWE-Bench Verified. Second, we filtered to include only repositories with permissive open-source licenses. Finally, recognizing that a subset of repositories might be GPU-dependent, tutorial-based, or contain minimal source code, we performed a content-based filter. We extracted the README files from all remaining candidates and employed an LLM-as-a-judge approach to automatically exclude repositories deemed unsuitable for training.

\textbf{Pull Request Filtering.}
We next extracted all pull requests merged into the ``main'' or ``master'' branches of the filtered repositories, resulting in a preliminary set of 6 million entries. To ensure quality, we utilized an LLM-as-a-judge to filter low-quality instances based on the available metadata: the git diff, the pull request description, and the merge commit message.

\textbf{Cheating Prevention.}
During the evaluation phase, it is imperative to prevent the model from exploiting git commands (e.g., \texttt{git log --all}) to access ground truth solutions~\citep{xiao2026mimo}. To address this, we execute a sanitization script immediately after initializing the task environment. This process, systematically removes metadata that follows the task's parent commit. It performs a hard reset, deletes all remote references and tags, and purges internal git files (e.g., \texttt{logs/}, \texttt{packed-refs}, and various \texttt{HEAD} files), thereby eliminating any trace of future solution history.
\begin{figure*}[t!] 
  \begin{center}
    \centerline{\includegraphics[width=0.9\textwidth]{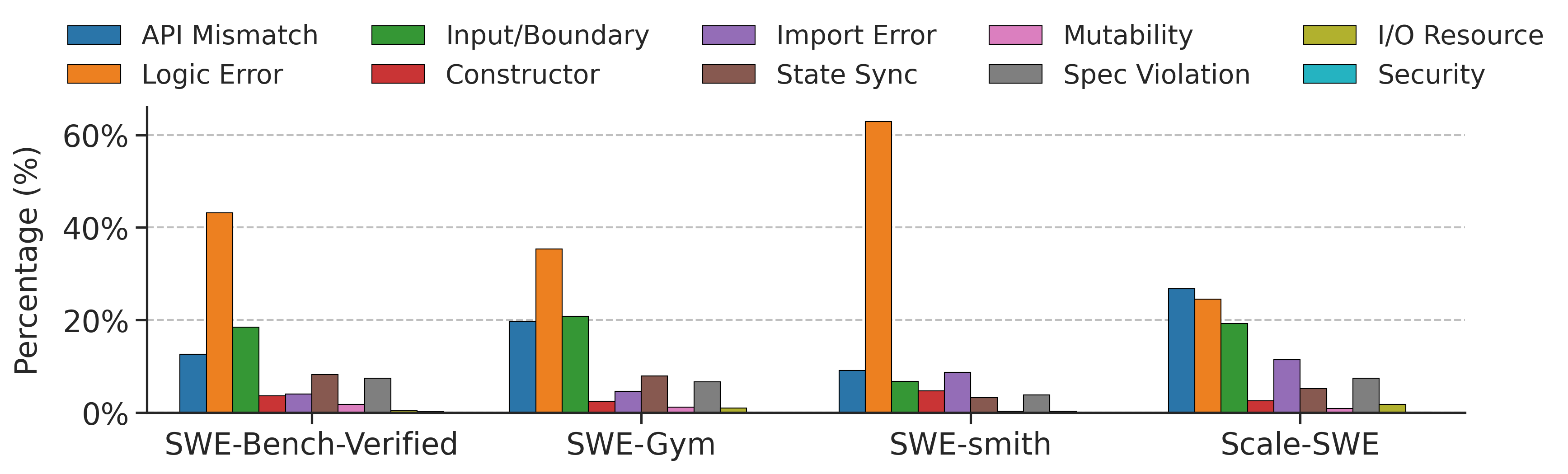}} 
    \caption{
Distribution of bug categories across different datasets. The bar chart compares the percentage of ten bug types within SWE-bench Verified, SWE-Gym, SWE-smith, and Scale-SWE. The categories are defined as: API Mismatch (incompatible signatures or parameter errors); Logic Error (flawed conditionals or control flow); Input/Boundary (edge case mishandling or validation failures); Constructor (object initialization errors); Import Error (missing modules or undefined symbols); State Sync (inconsistent internal state); Mutability (unintended side effects); Spec Violation (non-compliance with protocols); I/O Resource (file system or stream errors); and Security (improper scoping or access control).
    }
    \label{bug_distribution}
  \end{center}
  \vspace{-0.6cm} 
\end{figure*}
\subsection{Dataset Quality Assurance}

\textbf{Rule-based Check.} We implemented a rigorous filtering protocol based on test outcomes. We retained only those instances where: (1) all P2P tests pass and F2P tests fail on the original buggy codebase; and (2) all P2P and F2P tests pass upon application of the golden patch.

\textbf{Expert Check.} We engaged four senior Ph.D. students to manually audit 100 randomly sampled instances employing a cross-validation protocol. The audit confirmed that 94 instances were valid, featuring correct environments, unit tests, and problem statements. We attribute this high quality to three key factors: (1) We leveraged state-of-the-art models (e.g., Gemini 3 Pro) to synthesize accurate problem statements; (2) We employed an LLM-as-a-Judge approach to preemptively discard low-quality repositories and pull requests; and (3) We enforced a fixed execution order for P2P and F2P tests to prevent test pollution, where varying execution orders could otherwise lead to inconsistent results due to environment contamination.

\subsection{Dataset Statistics and Trajectory Distillation}

Ultimately, our agent-based construction pipeline yielded 100k successfully verified instances—the largest verified SWE benchmark to date. Leveraging our extensive collection of GitHub repositories, the resulting dataset achieves substantially greater repository diversity than existing SWE benchmarks, as detailed in Table~\ref{tab:dataset_comparison}. 
Unlike previous datasets that typically rely on synthetic generation or limited real-world sources, Scale-SWE is constructed entirely from 5.2k real repositories. This represents a 50\% increase in repository count over the next largest real-world benchmark (SWE-rebench with 3.5k repositories). This scale and diversity ensure that our benchmark better reflects the complexity and variety of real-world software engineering tasks.

The statistical characteristics of  Scale-SWE are presented in Table~\ref{tab:dataset_stat}. The dataset presents non-trivial software engineering challenges. The median instance requires modifications to at least 3 files and the addition of 43 lines of code, reflecting the task complexity. For evaluation robustness, instances contain an average of over 200 Pass-to-Pass (P2P) tests (median: 68), providing strong protection against regression, while an average of 5.69 Fail-to-Pass (F2P) tests per instance validates whether the LLM successfully resolves the specific issue.

To construct the training corpus, we distilled agent trajectories from a subset of 25k instances using the high-performance expert model DeepSeek-V3.2. For each instance, we conducted five independent sampling trials with a temperature of 0.95 and a maximum budget of 100 interaction turns. A trajectory was considered valid only if it culminated in a submission that passed all unit tests. This pipeline produced 71k high-quality trajectories totaling approximately 3.5 billion tokens. To ensure a fair comparison, we applied the identical pipeline to collect trajectories for SWE-Smith and SWE-Gym.

Figure \ref{bug_distribution} highlights the superior diversity of Scale-SWE. Synthetic datasets like SWE-smith exhibit a strong bias towards Logic Errors, failing to capture the intricacies of API Mismatches or State Synchronization issues common in large codebases. Similarly, SWE-Gym, despite being real-world sourced, suffers from low variety due to its restriction to only 11 repositories.

Conversely, Scale-SWE achieves a highly balanced distribution across all ten bug categories. By leveraging 5.2k real-world repositories and an automated environment-building pipeline, Scale-SWE effectively captures a wide array of defect types—from Constructor errors to Security flaws. This validates that scaling the source repositories and automating the execution pipeline are critical for synthesizing training data that faithfully represents real-world software evolution.

We adopt the bug taxonomy proposed in BugPilot~\citep{sonwane2025bugpilot} and employ DeepSeek v3.2 to automatically annotate the training instances.
\begin{table*}[h]
    \centering
    \caption{Comparison of Scale-SWE with existing software engineering (SWE) benchmarks. We contrast datasets across key dimensions: the number of executable instances—defined as those equipped with a Dockerized environment and validated via Fail-to-Pass (F2P) and Pass-to-Pass (P2P) tests—primary data source, repository diversity, and trajectory count.}
    \label{tab:dataset_comparison}
    \vspace{5pt}
    \begin{tabular}{lcccc} 
        \toprule
        \textbf{Dataset} & \textbf{Exec. instances} & \textbf{Primary Source} & \textbf{Repo.} & \textbf{Traj.}\\
        \midrule
        R2E-Gym~\citep{jain2025r2e}    & 4.6k & Synthetic & 10 & 3.3k\\
        SWE-Gym~\citep{pan2024training}    & 2.4k & Real & 11 & 491 \\
        SWE-smith~\citep{yang2025swe}  & 50k & Synthetic  & 128 & 5k \\
        SWE-Mirror~\citep{wang2025swe}  & 60k & Synthetic   & 40 & 12k\\
        SWE-rebench~\citep{badertdinov2025swe}  & 7.5k & Real  & 3.5k & N/A \\
        Scale-SWE  & \textbf{100k} & Real    & \textbf{5.2k} & \textbf{71k} \\
        \bottomrule
    \end{tabular}
\end{table*}

\begin{table*}[h]
    \centering
    \caption{Performance comparison on SWE-bench Verified. We categorize models into proprietary systems, open-source methods, and size-matched baselines.}
    \label{tab:swe_bench_results}
    \vspace{10pt}
    \begin{tabular}{llc}
        \toprule
        \textbf{Models} & \textbf{Base Model} & \textbf{SWE-bench (V)} \\
        \midrule
        
        \rowcolor{gray!15} 
        \multicolumn{3}{c}{\textit{Proprietary Models}} \\
        
        GPT-5.2 Thinking~\citep{openai2025gpt52} & - & \textbf{80.0} \\
        Claude Sonnet 4.5~\citep{anthropic2026sonnet45} & - & 77.2 \\
        Gemini 3 Pro~\citep{googledeepmind2026gemini3pro} & - & 76.2 \\
        MiniMax-M2.1~\citep{minimax2026m21} & - & 74.0 \\
        GLM-4.7~\citep{zhipu2026glm47} & - & 73.8 \\
        DeepSeek-V3.2~\citep{liu2025deepseek} & - & 73.1 \\
        Kimi K2 Thinking~\citep{team2025kimi} & - & 71.3 \\
        
        \midrule
        \rowcolor{gray!15}
        \multicolumn{3}{c}{\textit{Open Source Methods}} \\
        
        SWE-Gym-32B~\citep{pan2024training} & Qwen-2.5 coder &  20.6 \\
        SWE-Fixer-72B~\citep{xie2025swe} & Qwen2.5-72B & 32.8 \\
        R2E-Gym-32B~\citep{jain2025r2e} & Qwen-2.5-Coder & 34.4 \\
        SWE-rebench-72B~\citep{golubev2025training} & Qwen2.5-72B-Instruct & 39.0 \\
        SWE-smith-32B~\citep{yang2025swe} & Qwen2.5-32B & 40.2 \\
        SWE-RL~\citep{wei2025swe} & Llama3-70B &  41.0 \\
        Skywork-SWE-32B~\citep{zeng2025skywork} & Qwen2.5-Coder-32B-Instruct &  47.9 \\
        SWE-Mirror-LM-32B~\citep{wang2025swe} & Qwen2.5-Coder-32B-Instruct &  52.2 \\
        SWE-Lego-32B~\citep{tao2026swe} & Qwen3-32B &  52.6 \\
        KAT-Dev-32B~\citep{zhan2025kat} & - & \textbf{62.4} \\
        
        \midrule
        \rowcolor{gray!15}
        \multicolumn{3}{c}{\textit{Models of the same size}} \\
        
        Qwen3-30B-A3B-Instruct~\citep{qwen3technicalreport} & - &  22.0 \\
        Qwen3-Coder-30B-A3B-Instruct~\citep{qwen3technicalreport} & - &  51.6 \\
        GLM-4.7-Flash-30A3B~\citep{zhipu2026glm47flash} & - & \textbf{59.2} \\
        
        \midrule
        \rowcolor{gray!15}
        \multicolumn{3}{c}{\textit{Our Model}} \\
        
        \textbf{Scale-SWE-Agent} & Qwen3-30B-A3B-Instruct &  \textbf{64.0} \\
        \bottomrule
    \end{tabular}
\end{table*}

\section{Experiments}
\subsection{Experiment Setup}


\textbf{Agent Scaffolding.} We employed OpenHands~\citep{wang2025openhands}, an open-source, event-driven platform, as the unified agent framework for all experiments. OpenHands facilitates LLM agents to iteratively edit files, execute shell commands, and browse the web within sandboxed containers. We selected this framework due to its proven ability to establish robust and reproducible baselines on benchmarks such as SWE-Bench.

\textbf{Agent Post-training.} We perform post-training on the Qwen3-30B-A3B-Instruct~\citep{qwen3technicalreport} base model. The training process is configured with a learning rate of 1e-5, a batch size of 128, and a warmup ratio of 0.05, supporting a maximum context length of 131,072.

\textbf{Evaluation Benchmarks and Metrics}
Our evaluation is conducted on SWE-bench Verified~\citep{chowdhury2024introducing}, a benchmark comprising 500 high-quality, human-curated Python software issues. We report the Resolved Rate (\%), representing the proportion of instances for which the model generates a correct solution. Notably, although the models were trained with a sequence length of 131,072, we extended the context limit to 262,144 during inference to handle larger inputs.

\subsection{Experiment Results}
As shown in Table \ref{tab:swe_bench_results}, the Scale-SWE Agent demonstrates superior performance on the SWE-bench Verified.
First, regarding the impact of our scaling strategy, Scale-SWE Agent achieves a remarkable 42.0\% absolute improvement over its base model, Qwen3-30B-A3B-Instruct, boosting the pass rate from 22.0\% to 64.0\%.
Second, in comparison to models of the same size, our method significantly outperforms strong competitors, including Qwen3-Coder (51.6\%) and GLM-4.7-Flash (59.2\%).
Furthermore, Scale-SWE Agent exhibits exceptional efficiency, surpassing models with significantly larger parameter counts, such as SWE-RL (Llama3-70B) and SWE-Fixer-72B. Notably, it also exceeds the previous state-of-the-art open-source method, KAT-Dev-32B (62.4\%), and outperforms recent specialist models like SWE-Mirror and SWE-Lego by a margin of over 11\%. These results validate the effectiveness of scaling up SWE-style data for enhancing software engineering capabilities.

\begin{figure}[t!] 
  \begin{center}
    \centerline{\includegraphics[width=0.7\columnwidth]{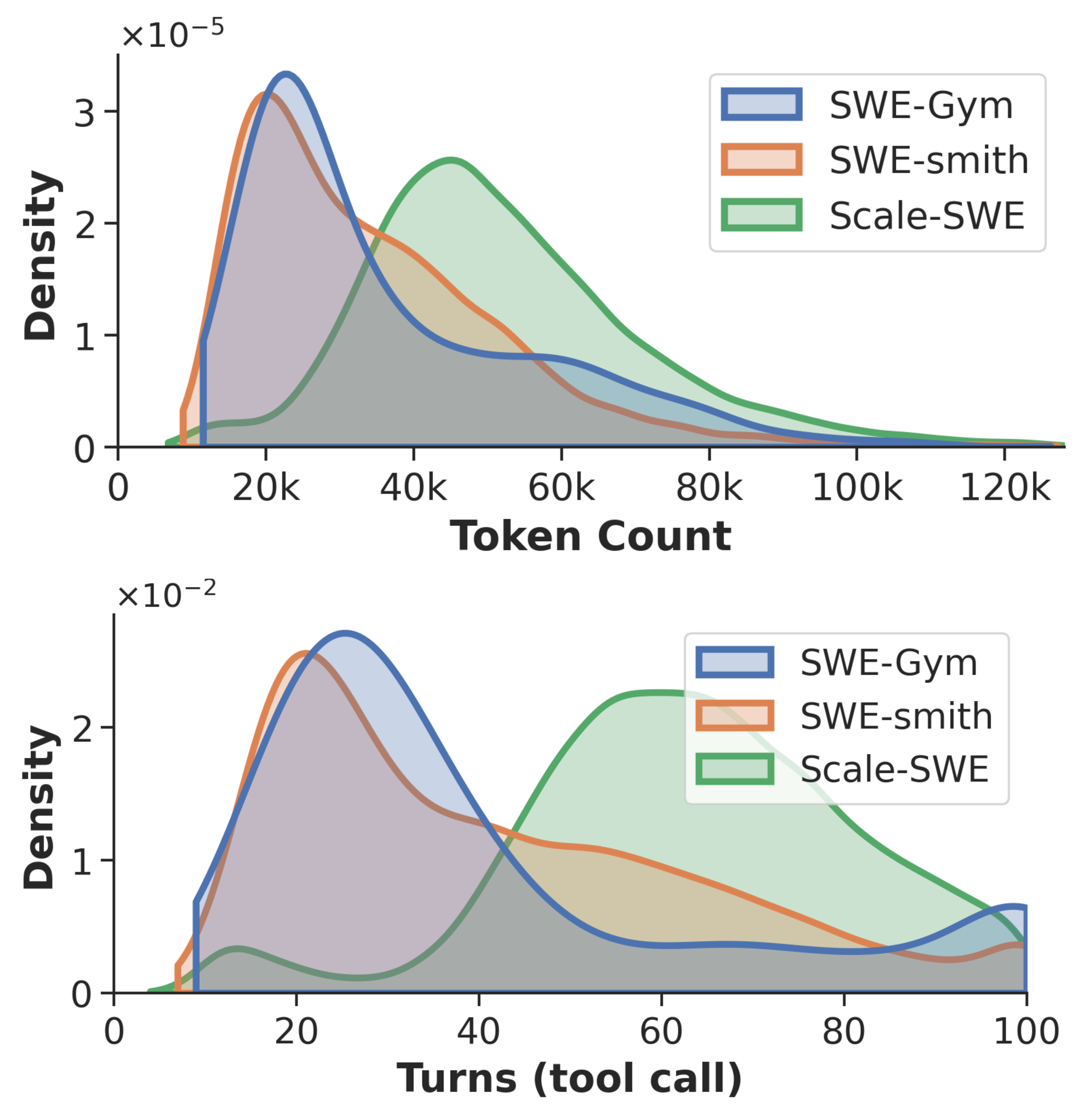}} 
    \caption{
        Comparison of distillation data statistics across different datasets. We show the probability density functions for (top) total token count and (bottom) the number of tool-call turns.
    }
    \label{turn_stat}
  \end{center}
  \vspace{-0.6cm} 
\end{figure}

To evaluate the efficacy of our data compared to existing alternatives, we conducted controlled experiments by performing distillation and SFT on SWE-Gym and SWE-smith using an identical pipeline. 
As shown in Table \ref{tab:swe_bench_baseline}, Scale-SWE significantly outperforms both baselines. Notably, despite SWE-smith possessing a considerably larger volume of instances than SWE-Gym, it yields slightly inferior performance. This performance gap suggests a diminishing return on purely synthetic data and underscores a critical insight: high-fidelity, real-world data is inherently more effective than massive-scale synthetic alternatives. This finding reinforces the importance of our large-scale ``real-data'' construction approach.

The distributions of interaction turns and token counts are presented in Figure \ref{turn_stat}. As illustrated in these figures, Scale-SWE tasks necessitate a greater number of turns for repository exploration and iterative debugging. This observation underscores the high complexity and difficulty inherent in the Scale-SWE dataset.

\begin{table}[h]
    \centering
    \caption{SFT performance comparison on SWE-bench Verified. All models are fine-tuned using the same distillation pipeline to ensure a fair comparison.}\label{tab:swe_bench_baseline}
    \vspace{5pt}
    \begin{tabular}{lc} 
        \toprule
        \textbf{Dataset Name} & \textbf{SWE-bench Verified} \\
        \midrule
        SWE-Gym & 54.8 \\
        SWE-smith & 54.6 \\
        Scale-SWE & \textbf{64.0} \\
        \bottomrule
    \end{tabular}
\end{table}

\section{Related Work}
\textbf{SWE Benchmark.} Since the introduction of the prevailing software engineering benchmark, SWE-bench~\citep{jimenez2023swe} and SWE-bench-Verified~\citep{chowdhury2024introducing}, many other benchmarks have emerged to assess multi-modal~\citep{yang2024swe}, multi-language~\citep{zan2025multi,rashid2025swe,guo2025omnigirl}, and long-horizon capabilities~\citep{deng2025swe}. These new benchmarks also evaluate whole repository generation~\citep{ding2025nl2repo}, scientific domain knowledge~\citep{duston2025ainsteinbench}, and other specialized abilities~\citep{ma2025swe,shetty2025gso}. Collectively, these benchmarks constitute a comprehensive evaluation ecosystem, establishing rigorous standards that assess the multifaceted capabilities required for autonomous software engineering.

\textbf{SWE Datasets.} High-quality data is pivotal for enhancing the programming capabilities of Large Language Models (LLMs). Recently, there has been a surge in repository-level software engineering datasets aimed at addressing complex coding tasks. Efforts to scale up SWE task instances generally fall into two categories. One line of work, including R2E-Gym~\citep{jain2025r2e}, SWE-smith~\citep{yang2025swe}, and SWE-Mirror~\citep{wang2025swe}, attempts to scale up training data through synthetic generation. Conversely, other works focus on mining real-world issues; for instance, SWE-Gym~\citep{pan2024training} constructed 2,400 executable instances restricted to 11 repositories, while SWE-rebench~\citep{badertdinov2025swe} further expanded this collection to 7,500 executable instances.

\textbf{SWE Models and Agents.} Recent advancements have introduced powerful models specialized for SWE tasks, including SWE-RL~\citep{wei2025swe}, SWE-Swiss~\citep{SWESwiss2025}, Kimi-Dev~\citep{yang2025kimi}, and KAT-Coder~\citep{zhan2025kat}. In parallel, frameworks such as SWE-agent~\citep{yang2024sweagent}, Mini-SWE-Agent~\citep{yang2024sweagent}, OpenHands~\citep{wang2025openhands}, and MOpenHands~\citep{zan2025multiswebench} serve as effective scaffolds to streamline interactions with development environments.

\section{Conclusion} In this work, we introduced Scale-SWE, a sandboxed multi‑agent framework that automates the construction of large‑scale, high‑quality software engineering data along with executable environments. By orchestrating specialized agents for environment setup, unit‑test generation, and task description synthesis, we processed six million real‑world pull requests to produce {Scale-SWE-Data}—a dataset of 100,000 instances that surpasses existing datasets in both scale and repository diversity. 
We further demonstrated the practical value of the dataset by distilling high‑quality trajectories and fine‑tuning Qwen3‑30B-A3B-Instruct to create {Scale-SWE-Agent}, which achieves substantial improvements on SWE‑Bench‑Verified—increasing the resolve rate from 22\% to 64\%. This advancement underscores the quality and utility of our data for training more capable code agents. 
We believe {Scale-SWE} opens new avenues for scalable SWE‑style dataset construction and provides a rich, open‑access resource to support the development of more capable LLM‑based software engineering agents. In future work, we aim to not only increase the volume of training data to fully leverage the potential of Scale-SWE-Data but also significantly broaden its linguistic scope. Specifically, we plan to extend our pipeline to support other major programming languages—such as Java, C, C++, and Rust—thereby fostering the development of truly language-agnostic software engineering agents.

\appendix
\newpage

\section{Scale-SWE workflow details.}\label{detail_workflow}
\subsection{Scale-SWE workflow Overview.}\label{sec:data}
\begin{figure*}[h]  
  \begin{center}
    \centerline{\includegraphics[width=\textwidth]{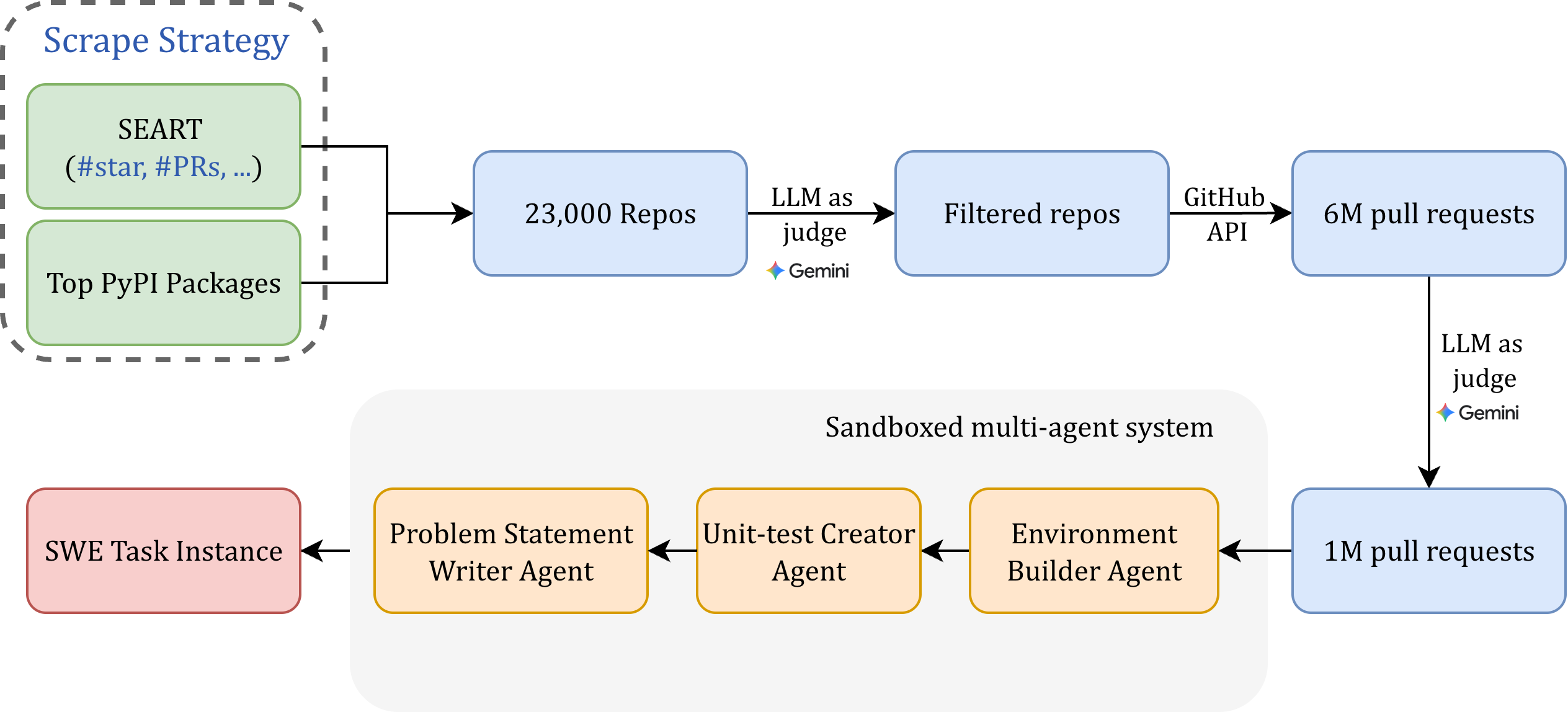}} 
    \caption{
Schematic workflow for automated Scale-SWE task synthesis. From an initial pool of 23k repositories and 6M pull requests, the pipeline utilizes LLM-as-a-judge to filter for quality and relevance. The selected 1M pull requests are then transformed into formal software engineering task instances via a sandboxed orchestration of specialized agents responsible for environment building, test creation, and statement writing.
    }\label{workflow_overview}
  \end{center}
\end{figure*}



\section{Scale-SWE Task Instance Structure}\label{task_instance}

The structure of a Scale-SWE task instance closely adheres to the SWE-bench standard, with a necessary adaptation to accommodate synthetic data. Specifically, since original developer-written ``fail-to-pass'' (F2P) tests are not available for all mined instances, we include a field for F2P scripts generated by our unit-test creator agent. A Scale-SWE task instance consists of the following fields:

\begin{itemize}
    \setlength{\itemsep}{0pt} 
    \item \textbf{instance\_id}: A unique identifier formatted as \texttt{\{user\}\_\{repo\}\_pr\{id\}}.
    \item \textbf{user}: The owner of the GitHub repository.
    \item \textbf{repo}: The name of the GitHub repository.
    \item \textbf{language}: The programming language of the codebase (currently Python).
    \item \textbf{workdir}: The working directory path within the environment.
    \item \textbf{image\_url}: The URL of the pre-built Docker image for the task.
    \item \textbf{patch}: The ground-truth patch (Golden Patch) from the corresponding pull request.
    \item \textbf{pr\_commit}: The commit hash of the pull request.
    \item \textbf{parent\_commit}: The commit hash of the parent commit (base state).
    \item \textbf{problem\_statement}: The issue description conveying the bug, provided to the model as input.
    \item \textbf{f2p\_patch}: The developer-written test patch containing tests that fail before the fix (if available).
    \item \textbf{f2p\_script}: The synthetic reproduction script generated by our unit-test creator agent (used when \texttt{f2p\_patch} is absent).
    \item \textbf{FAIL\_TO\_PASS}: A list of unit tests that fail when applied to the buggy version but pass after the fix.
    \item \textbf{PASS\_TO\_PASS}: A list of unit tests that pass in both the buggy and fixed versions (regression tests).
    \item \textbf{github\_url}: The URL of the original GitHub repository.
    \item \textbf{pre\_commands}: Commands executed immediately upon entering the container to revert future commit information and prevent data leakage.
\end{itemize}

\section{Implementation Details}\label{exp_detail}

The hyperparameters for SFT are detailed in Table \ref{tab:sft_param}.

\begin{table}[h]
    \centering
    \caption{Key hyperparameters in the SFT phase.}
    \label{tab:sft_param}
    \vspace{5pt}
    \begin{tabular}{lc} 
        \toprule
        \textbf{Hyperparameter} & \textbf{Value} \\
        \midrule
        Learning Rate & 1e-5 \\
        Base model & Qwen3-30B-A3B  \\
        Batch size & 128 \\
        Maximum Context Length & 131,072 \\
        Warmup ratio & 0.05 \\
        LR scheduler type & Cosine \\
        Epoch & 3 \\
        \bottomrule
    \end{tabular}
\end{table}



\section{Anti-hack Strategy}\label{anti_hack}
To ensure the integrity of the SWE-bench evaluation, we implement an anti-leak strategy to prevent the LLM from accessing ground truth solutions via Git history (e.g., using commands like \texttt{git log --all}). The following sanitization script is executed immediately after initializing the environment container:

\begin{lstlisting}[
    language=bash,               % 设定语言
    backgroundcolor=\color{bg},  % 背景颜色
    frame=tb,                    % 上下边框 (top, bottom)
    numbers=left,                % 行号显示在左侧
    numberstyle=\tiny\color{gray}, % 行号样式
    breaklines=true,             % 自动换行
    xleftmargin=15pt,            % 左边距
    basicstyle=\ttfamily\small,  % 代码字体
    keywordstyle=\color{blue}\bfseries, % 关键词高亮
    commentstyle=\color{green!40!black}, % 注释高亮
    stringstyle=\color{red},     % 字符串高亮
    framesep=5pt                 % 边框与内容的间距
]

git clean -fd -e '*.egg-info' -e '.tox' -e '.venv' && git checkout {parent_commit}
NEW_BRANCH="swe_bench_clean_main"
CURRENT_HEAD=$(git rev-parse HEAD)
git stash -a
git clean -fd
git reset --hard $CURRENT_HEAD
git stash pop || echo "No stash to apply or conflict occurred"

git config user.email "pre-agent@swalm.local" && git config user.name "Pre-Agent" && git add . && git commit -m "pre-agent commit"


CURRENT_3_PRIME=$(git rev-parse HEAD)

git for-each-ref --format="%(refname)" refs/remotes/ | xargs -I {} git update-ref -d {}
git tag -l | xargs -r git tag -d

rm -f .git/packed-refs
rm -f .git/ORIG_HEAD .git/FETCH_HEAD .git/MERGE_HEAD .git/CHERRY_PICK_HEAD .git/refs/stash
rm -rf .git/logs/

git update-ref refs/heads/${NEW_BRANCH} $CURRENT_3_PRIME
git symbolic-ref HEAD refs/heads/${NEW_BRANCH}

git for-each-ref --format="%(refname)" refs/heads/ | grep -v "refs/heads/${NEW_BRANCH}" | xargs -I {} git update-ref -d {}

git gc --prune=now --aggressive
\end{lstlisting}

\section{Prompts in Scale-SWE workflow}\label{prompts}

\begin{markdownbox}[title={Repository filtering prompt}]
You are an expert Software Architect and Engineer creating a high-quality benchmark dataset for evaluating autonomous software engineering agents (SWE-bench).

Your task is to analyze the provided README.md content and classify the repository as either **Suitable** or **Unsuitable** for benchmarking code generation and bug-fixing capabilities.

**Strict Negative Filters (Mark as 'Unsuitable' if ANY of these apply):**
1.  **Deep Learning & GPU Dependent:** Repositories focused on training/fine-tuning Neural Networks, LLMs, or using frameworks like PyTorch/TensorFlow for model training. Exclude anything requiring CUDA, GPUs, or specific pre-trained weights. We want *software logic*, not *model training scripts*.
2.  **API Wrappers & Thin Clients:** Repositories that primarily wrap external APIs (e.g., OpenAI SDKs, Telegram Bots, Cloud SDKs) or simply forward requests. If the code logic is just `client.call()`, it is invalid.
3.  **Non-Code Content:** "Awesome" lists, resource collections, interview questions, books, cheat sheets, or documentation-only sites.
4.  **Educational/Trivial:** Simple tutorials, "Hello World" examples, student homework, or demo apps with minimal complexity.

**Positive Criteria (Mark as 'Suitable' ONLY if the repository meets these standards):**
The repository must represent a substantial software engineering project with significant **internal logic**.
* **Examples:** Database engines, compilers/interpreters, complex utility libraries, backend web frameworks (implementation details, not just usage), algorithm libraries, network protocols, or file system implementations.
* The code should involve data structures, complex control flows, and purely algorithmic challenges independent of heavy external hardware environment.

**Input Handling:**
The README content below may be truncated. Make your decision based on the visible text. If the text clearly indicates a deep learning project or an API wrapper, reject it immediately.

You must respond with *only* one word: **Suitable** or **Unsuitable**

Here is the README.md content:
---
{readme_content}
---

\end{markdownbox}

\begin{markdownbox}[title={Pull request filtering prompt}]

You are an advanced classification assistant for screening Pull Requests (PRs).
Your task is to evaluate the PR diff I provide and determine if it is valuable for training an AI model's coding abilities.

You must respond with **only one word** in the following format exactly:
<Classification>

Where <Classification> is **only** one of "Suitable" or "Unsuitable".

"Unsuitable" (Useless) Criteria (Must be filtered out):
1.  **Non-Code Changes:**
    * Only modifications to documentation (e.g., README.md).
    * Only modifications to comments within the code.
2.  **Extremely Simple Code Changes:**
    * Only changing a parameter's default value, a constant, or a configuration item.
    * Only fixing typos in strings or comments.
    * Only code formatting changes (whitespace, indentation).
3.  **Auto-generated or Dependency Files:**
    * Only changes to lock files (package-lock.json, go.sum, etc.).
    * Only changes to auto-generated code or compiled artifacts.
    * Only changes to .gitignore or similar configuration files.

"Suitable" (Useful) Criteria:
* Includes substantive changes to code logic (e.g., fixing a bug, adding a feature, refactoring, optimization).
It is ok for Pull Request Description to be empty as long as the code is nontrivial.

Evaluate the following pull request conversation and PR diff, then provide your classification in the format <Classification> (only one word: "Suitable" or "Unsuitable").

**Pull Request Title:** {pr_title}
**Pull Request Description:** {pr_description}
**Merge Commit message:** {merge_commit_message}

**Diff:** (can be truncated if it is too long)
---
{pr_diff}
---

You can only answer
Suitable
or
Unsuitable

**only one word without any other words**

\end{markdownbox}

\begin{markdownbox}[title={Environment builder agent agent}]
Your goal is to clone, set up environment for a repository within the existing (base) conda environment so that all tests run successfully.
You are now at a GitHub repo at /workspace/{{repository}}.
The base image is Ubuntu 22.04 with Miniconda3 preinstalled you are already inside a conda shell (base environment).
You should install dependencies and configure everything using conda (and pip inside conda if necessary).

The repository is predominantly written in Python. Here are several tips for installing it:  
1. A good place to start is to look for a CONTRIBUTING.[md|rst] file, which will often contain instructions on how to install the repository and any dependencies it may have. Occasionally, the README.md file may also contain installation instructions.  
2. Usually, a repository may have setup.py or pyproject.toml files which can be used to install the package. pip install -e . is commonly used, although many packages will also require an additional specifier that installs development packages as well (e.g. pip install -e .[dev] or pip install -e .[tests]).  
3. To check whether the repository was installed successfully, run tests and see if they pass. You can usually find tests in a tests/ or test/ directory. You can run tests using pytest or unittest, depending on the framework used by the repository.
  **VERY IMPORTANT** YOU MUST APPEND ```--timeout=1800``` TO PYTEST, eg. "pytest ...  --timeout=1800".
4. Sometimes, you will need to install additional packages, often listed in a requirements.txt or environment.yml file. Also, be mindful of Ubuntu system dependencies that may need to be installed via apt-get (e.g. sudo apt-get install <package>).  
5. You MUST fix all errors encountered during testing warnings can be ignored.
6. YOU MUST DO ```pip freeze``` when all test cases pass. In this way I can get all exact version of all packages.
**IMPORTANT: YOU ARE NOT ALLOWED TO CHANGE ANY FILE TO FIX ENVIROMENT PROBLEM. INSTALL CORRECT VERSION PACKAGES TO FIX ENV INSTEAD OF CHANGE ANY FILE.**
Once you are finished with installing the repository, run the submit command to submit your changes for review.
\end{markdownbox}

\begin{markdownbox}[title={Unit-test creater agent prompt}]
You are an expert-level autonomous software engineer. Your **SOLE TASK** is to generate a single `pytest` test file named `fail_to_pass.py` to verify a Pull Request.
----------------------------------------------------------------
**CRITICAL OPERATIONAL RULES (READ FIRST)**
----------------------------------------------------------------
1. **NO SIMULATION:** Do NOT describe what a command *would* do. Do NOT invent terminal output.
2. **ACTION REQUIRED:** You are connected to a REAL terminal. To check a file, you MUST `cat` it. To run tests, you MUST `pytest` them.
3. **THOUGHT != ACTION:** Writing "I will run git show" in your thought process does nothing. You must output the actual code block/tool call to execute it.
4. **VERIFY REALITY:** Always inspect files (`ls`, `cat`) before assuming they exist.
----------------------------------------------------------------
**Provided Inputs**
----------------------------------------------------------------
**Repository Name:** {{repository}}
**Commit ID (Merge Commit):** {{commit_id}}
**Generated Problem Statement**: {{problem_statement}}
**PR Description in GitHub:** {{pr_description}}
**Commit Message:** {{commit_message}}
----------------------------------------------------------------
**MANDATED MULTI-PHASE PLAN**
----------------------------------------------------------------
You must follow this plan step-by-step.
------------------------------------------------
**PHASE 1: Analysis & Setup (Target State)**
------------------------------------------------
**Step 1: Inspect the changes**
- **EXECUTE** the following command immediately to see the real diff:
  `cd /workspace/{{repository}} && git show -m --first-parent --pretty=format: --patch {{commit_id}} > /workspace/diff.txt`
- **EXECUTE** `cat /workspace/diff.txt` to read the content.
- **Analysis:** Identify the high-level functions or classes that use the changed code.
------------------------------------------------
**PHASE 2: Test Generation (Iterative Writing)**
------------------------------------------------
**Step 2: Write the test file**
- **Goal:** Create `/workspace/{{repository}}/fail_to_pass.py` with **2 to 10 distinct test functions**.
- **Quantity Logic:** **The number of test cases you should generate should depend on the difficulty and extent of change of this commit.** (e.g., Use the lower end of the range for simple tweaks, and the higher end for complex logic overhauls).
- **Constraint:** ALL tests must FAIL on `{{commit_id}}^1` and PASS on `{{commit_id}}`.
- **Strategy for Diversity:**
    - **Vary Inputs:** Use different CLI arguments or config options.
    - **Vary Assertions:** Check for valid JSON structure, check for new keys (e.g., `check_result`), check specific values for `kconfig` vs `cmdline`.
- **CRITICAL ANTI-OVERFITTING RULE:**
    - Do NOT call new functions directly. Call the public API that invokes them.
- **ACTION:** **EXECUTE** the following command block to write the file (replace `...` with your python code):
  ```bash
  cat << 'EOF' > /workspace/{{repository}}/fail_to_pass.py
  import pytest
  import json
  # ... imports ...
  # ... Write 2-10 distinct test functions (quantity based on diff complexity) ...
  if __name__ == "__main__":
      sys.exit(pytest.main(["-v", __file__]))
  EOF
    ````
-----
## **PHASE 3: The "Time Travel" Verification (CRITICAL)**
You must prove your tests work by running them in the real environment.
**Step 3: Verify "After" State (Current HEAD)**
  - Ensure you are on `{{commit_id}}`.
  - **EXECUTE:** `pytest /workspace/{{repository}}/fail_to_pass.py`
  - **CHECK:** Do all tests PASS? If not, rewrite the file.
**Step 4: Verify "Before" State (Pre-PR)**
  - **EXECUTE:** `cd /workspace/{{repository}} && git checkout {{commit_id}}^1`
  - **EXECUTE:** `pytest /workspace/{{repository}}/fail_to_pass.py`
  - **CHECK:** Do all tests FAIL?
      - If they crash (ImportError), you failed the Anti-Overfitting Rule. **Rewrite.**
      - If they pass, you failed to reproduce the bug. **Rewrite.**
**Step 5: Return to HEAD**
  - **EXECUTE:** `cd /workspace/{{repository}} && git checkout {{commit_id}}`
  - If you had to rewrite tests in Step 4, repeat Step 3.
-----
## **PHASE 4: Final Submission**
**Step 6: Submit**
  - **EXECUTE:** `cat /workspace/{{repository}}/fail_to_pass.py` (to confirm final content).
  - **EXECUTE:** `submit`
\end{markdownbox}

\begin{markdownbox}[title={Problem statement writer agent prompt}]
You are an expert-level autonomous software engineer and open-source maintainer. Your **SOLE TASK** is to draft a concise, human-like GitHub Issue (Problem Statement) based on a provided Pull Request.
-----
## **CRITICAL OPERATIONAL RULES (READ FIRST)**
1.  **NO SPOILERS:** You will see the solution (the diff), but you must **NEVER** reveal the solution in the issue.
2.  **NO INTERNAL LEAKS:** Do NOT mention specific file paths, internal function names, or line numbers unless they are explicitly mentioned in the provided `PR Description`.
3.  **USER PERSPECTIVE:** Write as a user or developer stumbling upon the bug. Do not write as the person who just fixed it.
4.  **ACTION REQUIRED:** You are connected to a REAL terminal. You must execute commands to analyze the context before writing.
-----
## **Provided Inputs**
**Repository Name:** {{repository}}
**Commit ID (Merge Commit):** {{commit_id}}
**PR Description:** {{pr_description}}
**Commit Message:** {{commit_message}}
**F2P:** {{f2p}}
-----
## **MANDATED MULTI-PHASE PLAN**
You must follow this plan step-by-step.
-----
## **PHASE 1: Context & Diff Analysis**
**Step 1: Inspect the changes**
  - **Goal:** Understand what was broken by looking at how it was fixed.
  - **EXECUTE** the following command immediately to see the real diff:
    `cd /workspace/{{repository}} && git show -m --first-parent --pretty=format: --patch {{commit_id}} > /workspace/diff.txt`
  - **EXECUTE** `cat /workspace/diff.txt` to read the content.
  - **Analysis (Internal Monologue):**
    1.  Look at the code removed/changed in the diff.
    2.  Ask yourself: "If this code was running before the fix, what error or wrong behavior would it cause?"
    3.  Identify the public API or command that triggers this code path.
-----
## **PHASE 2: Reverse Engineering the Symptom**
**Step 2: Formulate the Bug Report Strategy**
  - **Constraint:** You generally know *why* it failed, but you must only describe *what* failed.
  - **Mental Check:**
      - Does the `PR Description` already describe the bug? If yes, align with it but refine clarity.
      - If the `PR Description` is empty or vague, use the `diff` to hallucinate the likely error message or wrong output based on logic.
-----
## **PHASE 3: Drafting the Issue**
**Step 3: Draft the content**
  - **Goal:** Create a natural, human-readable issue description.
  - **Guidelines (Strict adherence required):**
    1.  **Concise Title:** Choose a clear title describing the symptom (e.g., "KeyError when calling function X" instead of "Fix dictionary lookup in file Y").
    2.  **Reproduction Code:** Provide a "Minimal Reproducible Example".
          - It must look like a natural user script or snippet.
          - It must **NOT** be a unit test (no `assert` statements, no `self.assertEqual`).
          - It should strictly trigger the bug found in Phase 1.
    3.  **Expected vs Actual:**
          - **Actual:** Describe the error message (e.g., traceback) or the wrong data returned.
          - **Expected:** Describe what should have happened.
    4.  **Tone:** Casual but professional. Avoid excessive formatting.
    5.  **Secrecy:** Do not say "The bug is in line 50 of utils.py". Say "When I run this script, it crashes."
  - **ACTION:** **EXECUTE** the following command block to save your draft (replace `...` with your content). Ensure you wrap the final content in `[ISSUE]` tags.
<!-- end list -->
````bash
cat << 'EOF' > /workspace/issue_draft.txt
[ISSUE]
# [Title Here]
## Description
[Clear description of what you were trying to do and what went wrong]
## Reproduction Script
```python
# Provide a natural python snippet here that triggers the bug.
# Do NOT include assertions.
# Do NOT verify the fix here, just show how to break it.
````
## Actual Behavior
[Describe the error, traceback, or wrong output]
## Expected Behavior
[Describe what should have happened]
[/ISSUE]
EOF
```
------------------------------------------------
**PHASE 4: Final Verification & Submission**
------------------------------------------------
**Step 4: Safety Check**
- **EXECUTE:** `cat /workspace/issue_draft.txt`
- **Verification Questions:**
    1.  Did I mention a file name that is NOT in the PR description? -> *If yes, remove it.*
    2.  Did I explicitly explain the solution logic? -> *If yes, replace with symptom description.*
    3.  Is the reproduction script natural (no asserts)? -> *If no, rewrite it.*
**Step 5: Submit**
- **EXECUTE:** `submit`
\end{markdownbox}

\begin{markdownbox}[title={Bug type categorization prompt}]
# Task
Analyze the "Problem Statement" and "Golden Patch" (the fix) of a software issue to classify its root cause according to the provided taxonomy.

# Bug Taxonomy & Classification Criteria
* **A: API / signature mismatch or backward-compatibility break**
    - *Description:* Public interfaces change or fail to accept/forward expected parameters; options no longer propagated; removed/renamed methods.
    - *Signals:* TypeError for unexpected/unknown keyword, missing method attribute, inability to customize behavior that used to work.
* **B: Logic / conditional bug**
    - *Description:* Incorrect branching, inverted predicates, off-by-one comparisons, or misplaced conditions that alter behavior.
    - *Signals:* Wrong results for specific ranges/cases; behavior flips when a flag toggles; regression tied to a refactor of if/else logic.
* **C: Input validation, boundary, or sentinel handling error**
    - *Description:* Valid inputs rejected or invalid accepted; special values (NaN/None/NA/masked) mishandled due to comparison/identity semantics.
    - *Signals:* Edge cases fail while common cases pass; inconsistent behavior with empty inputs or special sentinels.
* **D: Incorrect argument forwarding, constructor, or inheritance contract break**
    - *Description:* Subclasses pass wrong args to super, fail to call base initializer, or expose mismatched signatures.
    - *Signals:* TypeError/AttributeError during object creation; missing base attributes; framework hooks not invoked.
* **E: Missing import / symbol / attribute error**
    - *Description:* Required names removed or not imported after refactor; attributes expected by callers no longer present.
    - *Signals:* NameError/AttributeError at runtime; module-level failures on import.
* **F: State consistency / bookkeeping / caching bug**
    - *Description:* Shared or stale state corrupts behavior across calls/instances; counters/heaps not updated; cache keys too coarse.
    - *Signals:* Nondeterministic results; memory growth; behavior depends on call order; leaked/stale entries.
* **G: Copy semantics, mutability aliasing, or in-place mutation of inputs**
    - *Description:* Wrong choice of shallow/deep copy; shared mutable defaults; functions mutate caller-provided objects.
    - *Signals:* Changes in one consumer affect another; unexpected side effects; duplicated or missing internal state.
* **H: Protocol / spec conformance bug**
    - *Description:* Behavior violates external specs (HTTP, OAuth, data interchange) or expected wire formats.
    - *Signals:* Clients reject responses; strict parsers fail; tests asserting spec rules break (e.g., HTTP HEAD body handling).
* **I: IO / filesystem / resource handling bug**
    - *Description:* Incorrect handling of paths/streams/resources; special-case short-circuits skip real writes; missing directory creation.
    - *Signals:* Truncated output; OSError/FileNotFoundError; behavior differs between stdout vs file.
* **J: Security / sensitive-data leakage due to logic oversight**
    - *Description:* Credentials/headers applied too broadly (e.g., to all domains) or without proper scoping/validation.
    - *Signals:* Tokens sent to unintended endpoints; security reviews flag over-permissive defaults.

# Input Data
**[Problem Statement]**
{ps_trunc}

**[Golden Patch]**
{patch_trunc}

# Instructions
1.  Carefully examine how the "Golden Patch" fixes the issue described in the "Problem Statement".
2.  Compare the fix logic against the "Description" and "Signals" in the Taxonomy.
3.  Provide a brief reasoning for your choice.
4.  Conclude with the category code inside tags, e.g., `<category>X</category>`.
\end{markdownbox}



\end{document}